\documentclass [twocolumn, aps, prl, superscriptaddress] {revtex4}
\usepackage{graphicx,latexsym}
\usepackage{color}

\begin{document}
\draft
\title {Giant Frictional Drag in Double Bilayer Graphene Heterostructures}
\author {Kayoung Lee}
\affiliation {Microelectronics Research Center, The University of Texas at Austin, Austin, TX 78758, USA}
\author {Jiamin Xue}
\affiliation {Microelectronics Research Center, The University of Texas at Austin, Austin, TX 78758, USA}
\author {David C. Dillen}
\affiliation {Microelectronics Research Center, The University of Texas at Austin, Austin, TX 78758, USA}
\author {Kenji Watanabe}
\affiliation{National Institute of Materials Science, 1-1 Namiki Tsukuba, Ibaraki 305-0044, Japan}
\author {Takashi Taniguchi}
\affiliation{National Institute of Materials Science, 1-1 Namiki Tsukuba, Ibaraki 305-0044, Japan}
\author {Emanuel Tutuc}
\affiliation {Microelectronics Research Center, The University of Texas at Austin, Austin, TX 78758, USA}
\date{\today}

\begin{abstract}
We study the frictional drag between carriers in two bilayer graphene flakes
separated by a 2 $-$ 5 nm thick hexagonal boron nitride dielectric.  At temperatures ($T$) lower than $\sim$ 10 K, we observe a large
anomalous negative drag emerging predominantly near the drag layer charge neutrality.
The anomalous drag resistivity increases dramatically with reducing {\it T}, and becomes comparable to the
layer resistivity at the lowest {\it T} = 1.5 K. At low $T$ the drag resistivity exhibits a breakdown of layer reciprocity.
A comparison of the drag resistivity and the drag layer Peltier coefficient suggests a thermoelectric origin of this anomalous drag.
\end{abstract}

\maketitle
Interaction between isolated electron systems in close proximity can produce a wealth of novel phenomena.
A particularly striking example is frictional drag where charge current ($I_{\rm Drive}$) flowing in one (drive)
layer induces a voltage drop in the opposite (drag) layer, $V_{\rm Drag} = R_{\rm D} I_{\rm Drive}$.
At the heart of the transresistance $R_{\rm D}$ are inter-layer couplings without particle exchange which can be mediated by e.g.,
momentum exchange~\cite{je1992}, energy transfer~\cite{js2012}, or phonons~\cite{hn1999}.
While being a sensitive probe of inter-layer interactions, the $R_{\rm D}$ values are generally much smaller than the layer resistance.
An exception occurs when the carriers in the two layers form a correlated state, yielding
$R_{\rm D}$ that can reach values comparable to the layer resistance. Indeed, this has been experimentally reported
in GaAs electron~\cite{mk2002}, or hole~\cite{et2004} double  layer systems,
in magnetic fields such that each layer has one half-filled Landau level~\cite{jp2004}.

Extensive experimental effort has been devoted to probe drag in electron-hole double layers,
using GaAs electron-hole double layers \cite{ac2008,js2009}, graphene double layers
\cite{sk2011,rg2012}, and most recently graphene-GaAs double layers \cite{ag2014}, motivated in
part by the search for equilibrium indirect exciton condensates. A common thread in these
experiments is an anomalous $R_D$ that increases with reducing $T$, along
with a breakdown of layer reciprocity when interchanging the drive and drag layers \cite{ac2008,js2009,ag2014}.
In this regard, double bilayer graphene separated by a thin hexagonal boron nitride (hBN) is a particularly compelling
system. The near parabolic energy-momentum dispersion in bilayer graphene allows the Coulomb to kinetic energy ratio
to be tuned via density, unlike monolayer graphene where this ratio is fixed
\cite{ap2013}. Moreover, the availability of ultra-thin dielectrics allows double
layers to be realized with interlayer spacing ($d$) down to a few nm, granting access to the strong coupling regime $d \ll l$, where $l$ is the inter-particle distance.
This effectively nests the two isolated electronic systems in the same plane.
Here, we investigate the frictional drag in double bilayer graphene heterostructures, consisting of two bilayer graphene
separated by a 2 $-$ 5 nm thick interlayer hBN dielectric, which allows us to explore the drag in a
wide range of layer densities and for all combinations of carrier polarity. Strikingly, we find a giant and negative
drag resistivity at charge neutrality, comparable to the layer resistivity at the lowest $T$.

The samples [Fig$.$ 1(a)] are fabricated using a layer-by-layer transfer process similar to samples discussed
in \cite{kl2014}.  The layer densities are tuned using a combination of back-gate ($V_{\rm BG}$),
and interlayer bias applied on the top bilayer ($V_{\rm TL}$)~\cite{sk2012}. The top ($\rho_{\rm T}$) and
bottom ($\rho_{\rm B}$) bilayer resistivities, as well as the frictional drag are probed using small
signal, low frequency lock-in techniques. We investigated five samples, labeled A-E, with
different interlayer spacing and layer mobilities. The key features of the drag data discussed below are similar in all
samples.

\begin{figure*}
\centering
\includegraphics[scale=0.7]{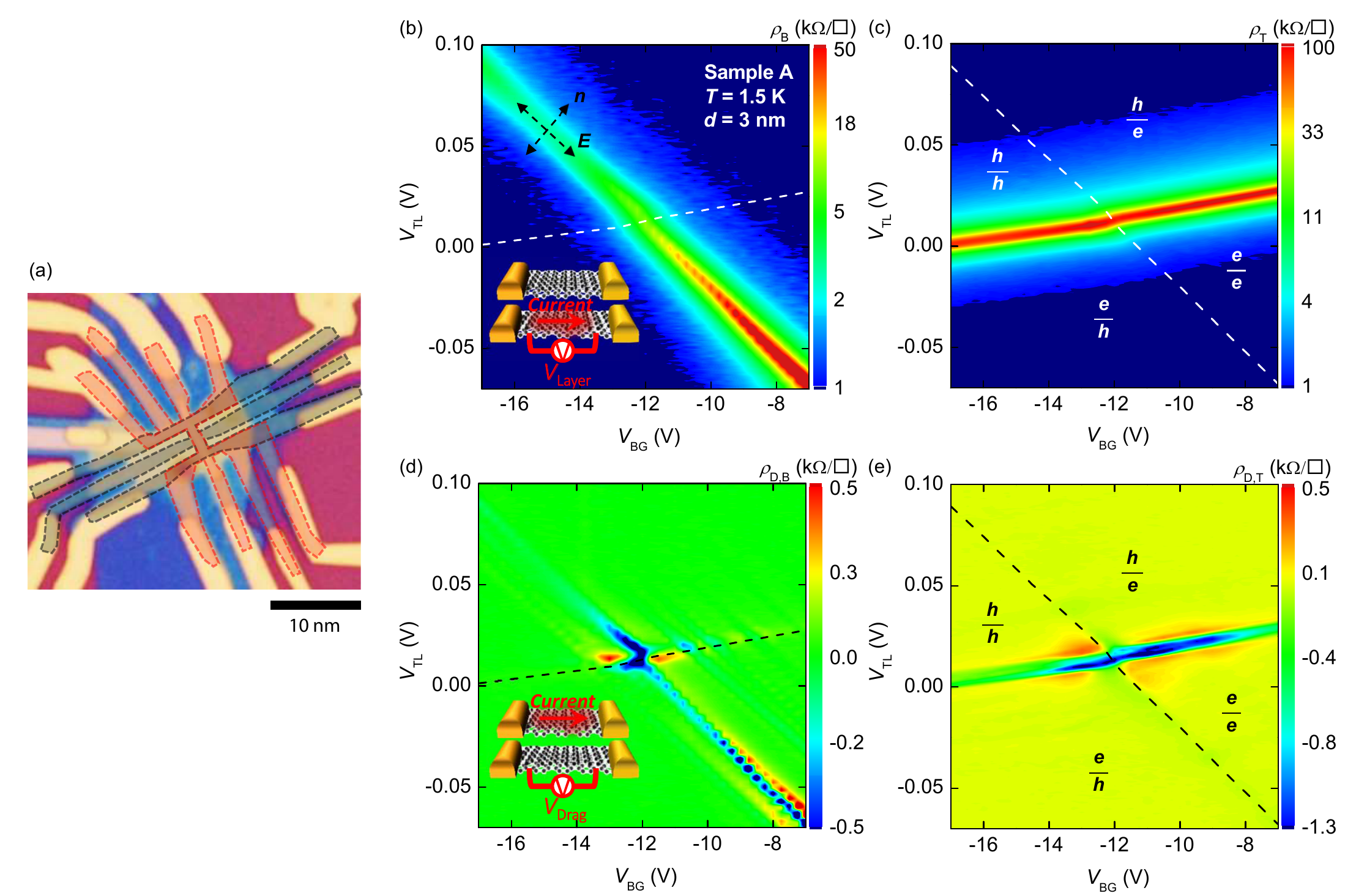}
\caption {\small{(a) Optical micrograph of a double bilayer graphene heterostructure.  The
red (gray) dashed contour lines mark the top (bottom) bilayer. (b) $\rho_{\rm B}$, and (c) $\rho_{\rm T}$ measured in Sample A as a function of
$V_{\rm BG}$ and $V_{\rm TL}$ at $T=1.5$ K. Panel (b) inset shows the sample and measurement schematic.
The white dashed lines in panels (b) and (c) mark the charge neutrality lines of the top and
bottom bilayers, respectively. Panel (c) shows the carrier type in the two bilayers in the four
quadrants defined by the two charge neutrality lines. (d) $\rho_{\rm D,B}$ and (e) $\rho_{\rm D,T}$
measured as a function of of $V_{\rm BG}$ and $V_{\rm TL}$ at $T=1.5$ K.}}
\label{fig1}
\end{figure*}

Figures 1(b) and 1(c) show $\rho_{\rm B}$ and $\rho_{\rm T}$ measured in Sample A at
$T = 1.5$ K. The bottom bilayer responds to $V_{\rm BG}$ and $V_{\rm TL}$ similar to a dual-gated
bilayer graphene, in which the density and transverse electric field ($E$) are controlled
independently \cite{kl2013}. The locus of high resistance points in Figs$.$ 1(b,c) marks the charge neutrality
lines for both bilayers. Figure 1(c) also shows the carrier type in each of the four quadrants
defined by the two charge neutrality lines. To examine variations in the drag resistance when
interchanging the drag and drive layers, we probe both the bottom ($\rho_{\rm D,B}$) and top
($\rho_{\rm D,T}$) drag resistivities, with the top or bottom bilayers serving as the drive
layers, respectively. Figures 1(d) and 1(e) show $\rho_{\rm D,B}$ and $\rho_{\rm D,T}$, respectively,
measured as a function of $V_{\rm BG}$ and $V_{\rm TL}$ in Sample A, at $T=1.5$ K. A comparison of Fig$.$
1(b,c) data on one hand, and Fig$.$ 1(d,e) data on the other, shows a large, negative
drag resistivity emerging predominantly near or at the drag layer charge neutrality.

\begin{figure}
\centering
\includegraphics[scale=1.05]{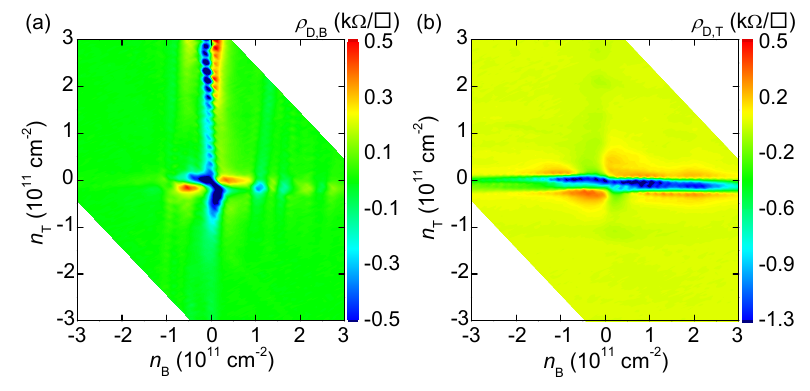}
\caption {\small{(a) $\rho_{\rm D,B}$ and (b) $\rho_{\rm D,T}$ as a function of $n_{\rm B}$ and $n_{\rm T}$,
measured at $T$ = 1.5 K. The data show a large drag resistivity
emerging along the drag layer charge neutrality, relatively insensitive to the drive layer
density.}}
\label{fig2}
\end{figure}

\begin{figure*}
\centering
\includegraphics[scale=0.63]{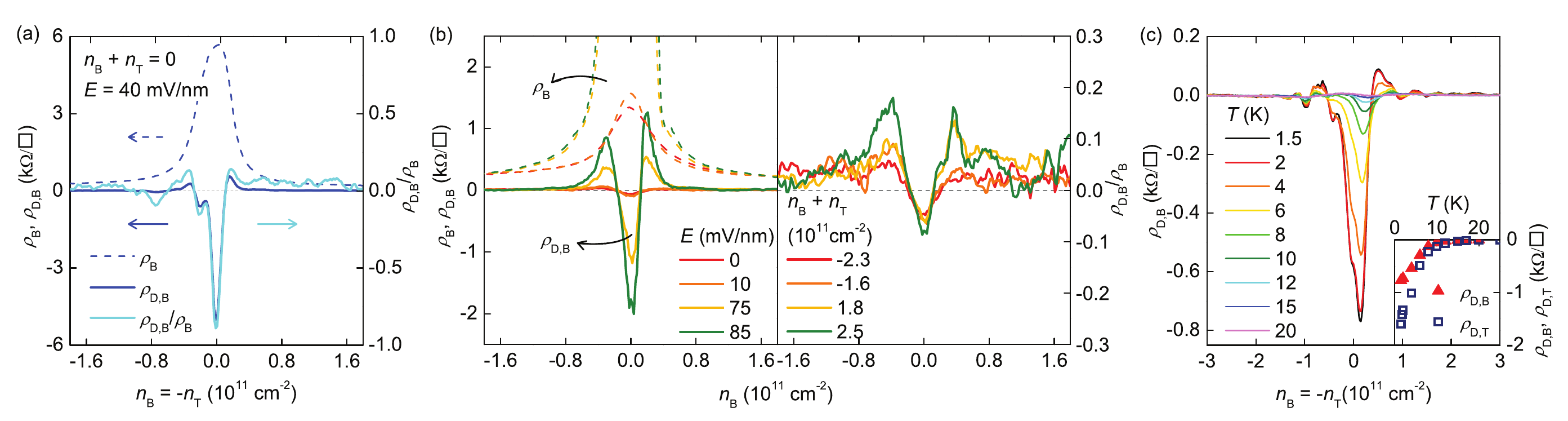}
\caption {\small{(a) $\rho_{\rm B}$, $\rho_{\rm D,B}$ (left axis), and $\rho_{\rm D,B}$/$\rho_{\rm B}$ (right axis) as a function of $n_{\rm B}$ =
$-n_{\rm T}$, measured at $T=1.5$ K in Sample A.  The $\rho_{\rm D,B}$ and $\rho_{\rm B}$ values are comparable at DNP. The
$E$-field across the bottom bilayer (drag layer) is 40 mV/nm at DNP. (b) Left panel: $\rho_{\rm B}$ (dashed lines),
and $\rho_{\rm D,B}$ (solid lines) vs. $n_{\rm B}$ in Sample A at different $E$ values in the bottom bilayer at $T=1.5$ K.
Right panel: $\rho_{\rm D,B}$/$\rho_{\rm B}$ vs. $n_{\rm B}$ corresponding to the left panel data. The data were acquired at
constant $n_{\rm B}+n_{\rm T}$ total density values. (c) $\rho_{\rm D,B}$ as a function of $n_{\rm B}$ = $-n_{\rm T}$,
in the proximity of DNP at different $T$, measured in Sample A in a separate cooldown. The inset shows  $\rho_{\rm D,B}$ and $\rho_{\rm D,T}$
vs. $T$ at the DNP.}}
\label{fig3}
\end{figure*}

To better visualize Fig$.$ 1(d,e) data, in Fig$.$ 2 we plot $\rho_{\rm D,B}$ [panel (a)] and
$\rho_{\rm D,T}$ [panel (b)] as a function of top ($n_{\rm T}$) and bottom ($n_{\rm B}$) bilayer densities, converted from
$V_{\rm BG}$ and $V_{\rm TL}$. The $n_{\rm T}$ and $n_{\rm B}$ values are related to the applied $V_{\rm BG}$ and $V_{\rm TL}$ biases,
referenced with respect to $n_{\rm B}=n_{\rm T}=0$, via:
$eV_{\rm BG}=e^2(n_{\rm B}+n_{\rm T})/C_{\rm BG}+\mu_{\rm B}$ and $eV_{\rm TL}=-e^2 n_{\rm T}/C_{\rm int}+\mu_{\rm B}-\mu_{\rm T}$,
where $C_{\rm BG}$ and $C_{\rm int}$ are the back-gate and interlayer capacitances, $\mu_{\rm T}$ and $\mu_{\rm B}$ are the top and bottom bilayers chemical
potentials, respectively, $e$ is the electron charge. To convert $V_{\rm BG}$ and $V_{\rm TL}$ to layer densities we use the density-dependent
chemical potential determined experimentally \cite{kl2014}. The $C_{\rm BG}$ and $C_{\rm int}$ values
are determined using magnetotransport measurements of individual bilayers \cite{bf2015}.
Figure 2 reveals a number of interesting features. First, $\rho_{\rm D,B}$ is large in the proximity of $n_{\rm B}$ = 0 line in Fig$.$ 2(a),
while $\rho_{\rm D,T}$ is large near $n_{\rm T}=0$ line in Fig$.$ 2(b). Near the double neutrality point (DNP), $n_{\rm B}$ = $n_{\rm T}$ = 0
$\rho_{\rm D,B}$ and $\rho_{\rm D,T}$ reach values close to 1 k$\Omega$. Second, the reciprocity with
respect to interchanging the drag and drive layers breaks down, i.e. $\rho_{\rm D,B}(n_{\rm B},n_{\rm T})\neq\rho_{\rm D,T}(n_{\rm B},n_{\rm T})$
in Fig$.$ 2.

In light of the anomalous drag observed in Figs$.$ 1 and 2, in the following we examine the drag layer resistivity in more detail,
concentrating on the drag layer density, and transverse electric field ($E$) dependencies. The latter is relevant for
bilayer graphene as the energy-momentum dispersion changes with $E$, concomitant with gap opening at charge neutrality \cite{em2006}.
Figure 3(a) shows Sample A $\rho_{\rm B}$, $\rho_{\rm D,B}$, and the corresponding normalized drag $\rho_{\rm D,B}/\rho_{\rm B}$ as a function
of $n_{\rm B}$ = $-n_{\rm T}$, namely at equal density in the two bilayers, with opposite polarity carriers.
$\rho_{\rm D,B}$ shows a very strong, negative peak at DNP, which surprisingly becomes comparable to $\rho_{\rm B}$
at $T=1.5$ K. As $n_{\rm B}$ = $-n_{\rm T}$ increases $\rho_{\rm D,B}$ changes sign, becomes positive at a finite $|n_{\rm B}|$, and then vanishes
as $|n_{\rm B}|$ increases further.

Figure 3(b) shows $\rho_{\rm B}$, $\rho_{\rm D,B}$ (left panel), and
$\rho_{\rm D,B}/\rho_{\rm B}$ (right panel) vs. $n_{\rm B}$ in the proximity of $n_{\rm B}$ = 0 and $n_{\rm T}$ $\neq$ 0.  The negative $\rho_{\rm D,B}$ at
$n_{\rm B}$ = 0 is notable, similar to the large, negative $\rho_{\rm D,B}$ peak at DNP in Fig$.$ 3(a).
However, the magnitude of $\rho_{\rm D,B}/\rho_{\rm B}$ at $n_{\rm B}$ = 0 and $n_{\rm T}$ $\neq$ 0 is smaller than that at DNP.
As $|n_{\rm B}|$ increases, $\rho_{\rm D,B}$ changes polarity, and becomes positive, consistent with the observed trend at DNP,
albeit with a lower magnitude. An examination of the electrostatics in double layers shows that at $n_{\rm B}$ = 0,
the $E$ value across the bottom bilayer changes as $n_{\rm T}$ changes as indicated in Fig$.$ 3(b) legend
(Supplementary Material).  We observe that $\rho_{\rm D,B}$ at $n_{\rm B}$ = 0 grows as $\rho_{\rm B}$ increases with
increasing $E$-field, leading to a relatively constant $\rho_{\rm D,B}/\rho_{\rm B}$ ratio.

Figure 3(c) shows $\rho_{\rm D,B}$ as a function of $n_{\rm B}$ = $-n_{\rm T}$ at different $T$ in Sample A, showing a large, negative
drag at DNP. We note that Fig$.$ 3(a,b) and Fig$.$ 3(c) were collected in separate cooldowns. Similar to Fig$.$ 3(a) data,
$\rho_{\rm D,B}$ becomes positive as $|n_{\rm B}|$ increases, and subsequently decreases towards zero with
increasing density. The inset of Fig$.$ 3(c) summarizes the $T$-dependence of the negative peak of both
$\rho_{\rm D,B}$ and $\rho_{\rm D,T}$ at DNP, showing a decrease of the drag resistivity with increasing $T$.
At the lowest $T$, mesoscopic fluctuations \cite{sk2012} are also noticeable in the proximity of DNP in Fig$.$ 3(c),
superimposed onto the large negative drag.

The experimental observations in Figs$.$ 1-3 have several anomalous features at variance with existing Coulomb drag theories.
It is tempting to interpret the giant drag that develops at DNP at low $T$ {\it prima facie}
as a signature of a correlated state of the two layers. However, the fact that the drag voltage is negative,
namely opposite to the electric field in the drive layer, coupled with the layer reciprocity breakdown casts doubt on this interpretation.
Moreover, the increasing $\rho_{\rm D}$ observed with decreasing $T$ [Fig. 3(c)] is opposite to the expected dependence for momentum transfer mediated drag \cite{je1992} .
The increasing drag at the lowest $T$, coupled with the apparent breakdown of reciprocity bears similarity with data reported in electron-hole double layers in GaAs-AlGaAs \cite{ac2008} or GaAs-graphene heterostructures \cite{ag2014}. We note that the interlayer separations in \cite{ac2008,ag2014} were larger than 10 nm, and the magnitude
of the measured drag resistivity was two orders of magnitude smaller than the values probed in the double bilayer graphene heterostructures investigated here.
Indeed, the $\rho_{D,B} \approx \rho_{B}$ is a dramatic signature of the strong coupling regime in double layers.

To gain insight into the origin of the anomalous drag we first note that the $\rho_{\rm D,B}$ and $\rho_{\rm B}$ peaks in Fig. 3(a) have similar widths.
The giant peak at the DNP is reminiscent of energy drag near charge neutrality in double monolayer graphene heterostructures \cite{js2012,rg2012},
where Coulomb mediated vertical energy transfer coupled with correlated density inhomogeneity in the two layers yields a drag resistivity of thermoelectric origin,
with the polarity determined by interlayer correlations $\langle\delta\mu_{\rm B}\delta\mu_{\rm T}\rangle$.
To assess the role of thermoelectricity in our measurements we use the Mott relation for the Peltier coefficient \cite{AM,yz2009}:
\begin{equation}
Q=\frac{\pi^2 k_B^2 T^2}{3e} \frac{\partial\sigma/\partial\mu}{\sigma}
\end{equation}
where $k_B$ is the Boltzmann constant, and $\sigma$ the layer conductivity. Using Eq$.$ (1) along with $\sigma=1/\rho_{\rm B}$
measured in the bottom bilayer graphene, the experimental $\mu$ vs. $n_{\rm B}$ data (Fig$.$ S1)
\cite{kl2014}, and $n_B$ vs. $V_{\rm BG}$ and $V_{\rm TL}$ (Fig. S2), we obtain $Q_{\rm B}$ vs. $\mu_{\rm B}$.

In Fig$.$ 4(a) (main panel) we compare the $\mu_{\rm B}$ dependence of $\rho_{\rm D,B}$ and $-\partial Q_{\rm B}$/$\partial\mu_B$ in Sample A at $T=1.5$ K.
Figure 4(a) inset shows $\rho_{\rm B}$ vs$.$ $\mu_{\rm B}$.  Both $\rho_{\rm D,B}$ and $\rho_{\rm B}$ were measured while sweeping
the layer densities such that $n_{\rm B}=-n_{\rm T}$. Remarkably, both $\rho_{\rm D,B}$ and $-\partial
Q_{\rm B}$/$\partial\mu_B$ show a peak at charge neutrality, change polarity as $|\mu_{\rm B}|$ increases,
and vanish at even larger $|\mu_{\rm B}|$ values. Interestingly, the peak structure of energy drag in
Ref$.$ \cite{js2012} arises from $\partial Q$/$\partial\mu$.

The striking similarity between the $\mu_{\rm B}$-dependence of $\rho_{\rm D,B}$
and $-\partial Q_{\rm B}$/$\partial\mu_B$ strongly suggests a thermoelectric origin for
the large frictional drag observed at low $T$ in our double bilayer graphene.
To further test this hypothesis, in Fig$.$ 4(b) we compare the $\mu$ value at which $\rho_{\rm D}$
changes polarity ($|\mu_{\rm Drag=0}|$), and the $\mu$ value at which the drag layer
$\partial Q$/$\partial\mu$ changes its polarity ($|\mu_{\rm dQ/d\mu=0}|$) for multiple samples.
The $|\mu_{\rm Drag=0}|$ and $|\mu_{\rm dQ/d\mu=0}|$ are averaged over the $\mu$ values on both
electron and hole branches, and represent the half width of the $\rho_{\rm D}$ peak and
the drag layer $\partial Q$/$\partial\mu$ peak, respectively. The $|\mu_{\rm Drag=0}|$
and $|\mu_{\rm dQ/d\mu=0}|$ values are determined using frictional drag measurements in either bottom or top bilayer graphene
from five samples with different interlayer thickness and layer mobility.  Furthermore, the data are
collected at different drive layer densities, not only at DNP.  Figure 4(b) clearly indicates that $|\mu_{\rm Drag=0}|$ agrees very
well with $|\mu_{\rm dQ/d\mu=0}|$, suggesting that the overall behavior of the anomalous
drag at low $T$ is governed by the drag layer $\partial Q$/$\partial\mu$.
Consistent with Figs$.$ 1 and 2 data showing that $\rho_{D}$ depends largely on the
drag layer density, we do not find a correlation between the drag resistivity and the
drive layer $\partial Q$/$\partial\mu$.

While reminiscent of energy drag, the giant drag measured here deviates from the simple energy drag picture presented in Ref.~\cite{js2012}.
Also striking is the layer non-reciprocity, amplified by the giant drag at DNP [Fig$.$ 3(a)]. We note that Ref.~\cite{js2012} assumes
fully overlapping layers with identical geometries, and contact configurations. In contrast, in the actual devices examined here
the geometry and contact configurations of the drive/drag layers are different [Fig$.$ 1(a)]. As a result, anisotropic heat flow
due to sample geometry~\cite{js2013} as well as Peltier heating outside of the active layers may contribute to the layer
non-reciprocity in our drag measurements. A fuller understanding of the origin of broken layer reciprocity at low $T$ is the subject of intense current research.

\begin{figure}
\centering
\includegraphics[scale=0.7]{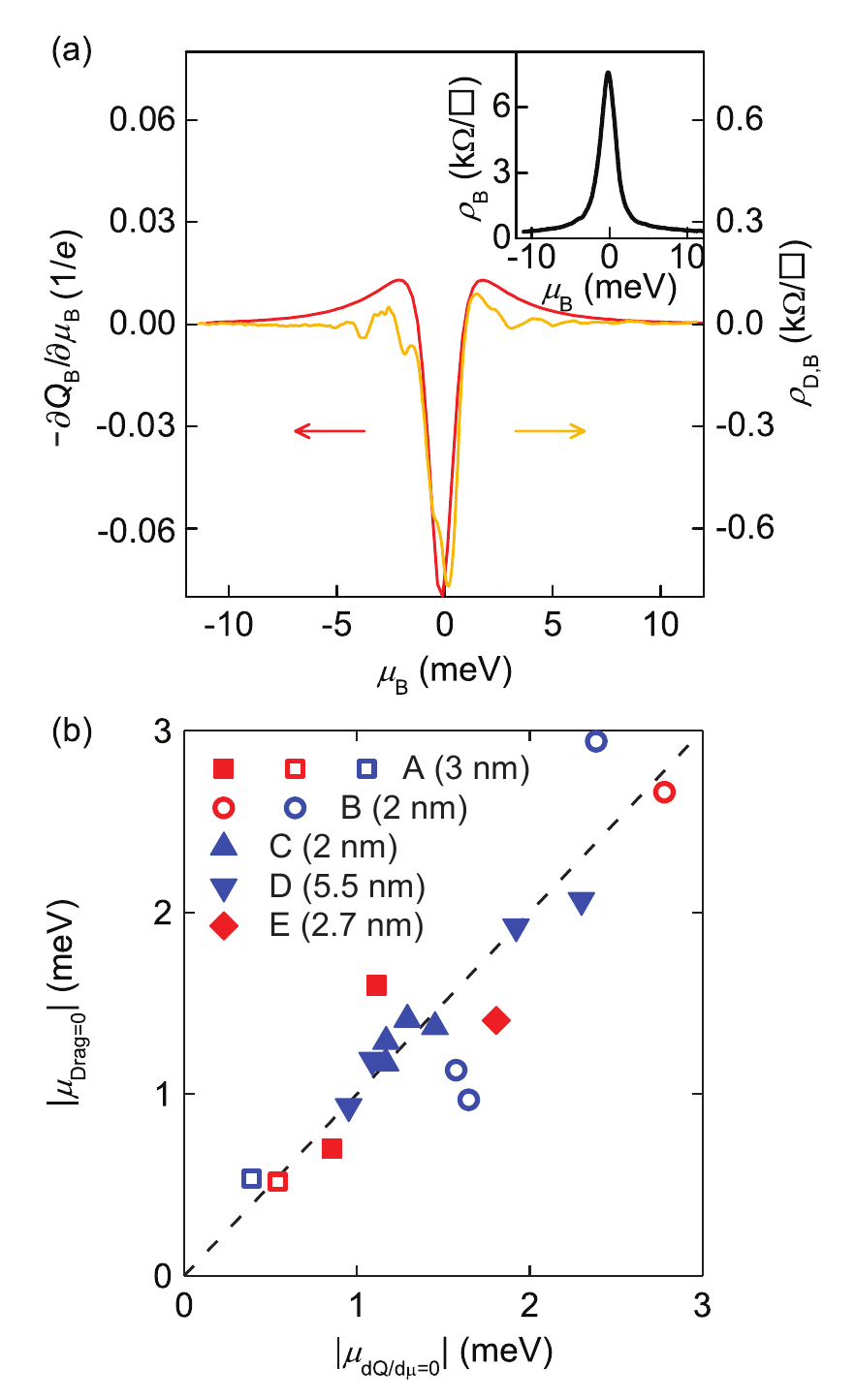}
\caption {\small{(a) $-\partial Q$/$\partial\mu_{\rm B}$ (red) and $\rho_{\rm D,B}$ (yellow) vs$.$
$\mu_{\rm B}$ in Sample A at $T=1.5$ K.  The inset shows the measured $\rho_{\rm B}$ vs. $\mu_{\rm B}$ data
used to calculate $\partial Q$/$\partial\mu_{\rm B}$. The data were acquired by sweeping the layer densities
such that $n_{\rm B}=-n_{\rm T}$. (b) $|\mu_{\rm Drag=0}|$ as a function of $|\mu_{\rm dQ/d\mu=0}|$ of the drag layer for five samples,
with different interlayer spacing shown in the legend. The open (closed)
symbols mark data measured using the top (bottom) bilayer as drag layer.
The red (blue) symbols represent data measured at zero (finite) drive layer density.}}
\label{fig4}
\end{figure}

The polarity of the energy drag is determined by the sign of potential fluctuations
in graphene, $\langle \delta \mu_{\rm B} \delta \mu_{\rm T} \rangle$ \cite{js2012}. A negative drag of thermoelectric origin
measured at DNP indicates that $\langle \delta \mu_{\rm B} \delta \mu_{\rm T} \rangle < 0$.
This suggests that strain \cite{mg2012}, rather than charged impurities \cite{jx2011} dominates the density inhomogeneity.
For impurity induced inhomogeneity $\langle\delta\mu_{\rm B}\delta\mu_{\rm T}\rangle > 0$, and a positive drag is expected at charge neutrality.
The clearly developed, broken symmetry integer quantum Hall states in our samples (Supplementary Material) also prove
the high sample quality with low level of impurities.

Lastly, we discuss similarities and differences between the energy drag previously observed in
double monolayer graphene heterostructures \cite{rg2012, mt2013}, and the drag in double bilayer
graphene heterostructures. The drag in monolayer graphene shows a peak at the DNP, has a positive value,
and is maximum at higher temperatures, $T \simeq 70$ K. The positive
drag at DNP is understood as energy drag where impurity induced disorder creates a positive
correlation of the layer chemical potential fluctuations $\langle\delta\mu_{\rm B}\delta\mu_{\rm T}\rangle$
\cite{js2012}. Interestingly, a comparison of the monolayer and bilayer graphene Peltier
coefficients using Eq$.$ (1) shows that the higher density of states and smaller $\sigma$
at charge neutrality in bilayer graphene yields a much larger $\partial Q$/$\partial\mu$,
and consequently larger drag at charge neutrality by comparison to monolayer graphene,
in agreement with the experimental observations (see Supplementary Material).

In summary, we report an anomalous giant, negative frictional drag $\simeq 1$ k$\Omega$ in high mobility double bilayer graphene near
the drag layer charge neutrality at temperatures lower than 10 K, with values approaching that of layer resistivity.
The drag increases with decreasing $T$ down to $T = 1.5\, {\rm K}$, and does not obey the layer reciprocity.
This opens an unanticipated playground for exploring new electron-interaction mediated phenomena in double layer systems even at zero field.

\begin{acknowledgments}
This work was supported by the SRC Nanoelectronics Research Initiative, and Intel Corp. We thank Justin Song for illuminating discussions,
and S$.$ Son for technical assistance.
\end{acknowledgments}

Note added: during the preparation of this manuscript we became aware of a related work described in Ref.~\cite{jl2016}.

\end{document}